\documentstyle[12pt,
         epsfig]{article}
\def\beq{\begin{equation}} 
\def\eeq{\end{equation}} 
\def\bea{\begin{eqnarray}} 
\def\eea{\end{eqnarray}}
\def\ra{\rightarrow}

\newcommand{\nc}{\newcommand}
\nc{\F}{{\cal F}}

\nc{\A}{{\cal A}}
\nc{\al}{\alpha} 
\nc{\be}{\beta}
\nc{\de}{\delta} 
\nc{\eps}{\epsilon}
\nc{\th}{\theta} 
\nc{\ups}{\upsilon}
\nc{\la}{\lambda} 
\nc{\si}{\sigma}
\nc{\PR}{{\bf Phys. Rev.~}}
\nc{\PL}{{\bf Phys. Lett.~}}
\nc{\NP}{{\bf Nucl. Phys.~}}
\nc{\PRL}{{\bf Phys. Rev. Lett.~}}
\begin{document}
\begin{titlepage}
\begin{flushright}
hep-ph/9811371{\hskip.5cm}\\
IOA-TH.12/98\\
NTUA 100/98
\end{flushright}
\begin{centering}
\vspace{.3in}
{\bf \bf Q-Balls and \\
the Proton Stability in Supersymmetric Theories. }\\
\vspace{2 cm}
{M. Axenides$^1$, E.G. Floratos$^1$, 
G.K. Leontaris$^2$ and N.D. Tracas$^3$} \\
\vskip 1cm
{\it {$^1$ Institute of Nuclear Physics, N.R.C.S. Democritos, 153 10 Athens, Greece}}\\
{\it {$^2$Physics Department, University of Ioannina,
Ioannina, GR 45 110, Greece}}\\
{\it {$^3$Physics Department, National Technical University, 
       Athens 157 73, Greece}}\\
\vspace{1.5cm}
{\bf Abstract}\\
\end{centering}
\vspace{.1in}
Abelian non-topological solitons with Baryon and/or Lepton 
quantum numbers naturally appear in the spectrum of the minimal
supersymmetric standard model. They arise  as a consequence of the existence
of flat directions in the potential lifted by non-renormalizable
operators and SUSY breaking. We examine the conditions that these
operators should satisfy in order to ensure proton stability and
present a realistic string model which fulfils these requirements. 
We further identify a generic $U(1)$ breaking term in the scalar
potential and discuss its effect of rendering Q-balls unstable.
\vfill
\hrule width 6.7cm 
\begin{flushleft}
November 1998
\end{flushleft}

 \end{titlepage}
\section{Introduction }

Non-topological solitons, abelian or nonabelian,
are finite energy configurations which appear
in the spectrum of field theories with global symmetries.
\cite{R,FLS,LP}.
They arise due to the
appearance of appropriate couplings in the scalar
potential that effectively cause a $Q$ number of scalar particles
of mass $m$ to form coherent bound states with binding energy $E/Q <m$. 
Although the general scaling property
of their total energy $ E \equiv Q^s (s<1)$ receives both
surface and volume contributions, there is
a special class of such configurations in the large $Q$ limit
with $s=1$ whose existence persists in the strict ``thermodynamic limit''
$ (V\ra \infty , E/Q\equiv const)$
\cite{C,SCA}. 

In the context of the minimal electroweak theory the presence of $B$ and $L$ balls
associated with the perturbatively conserved Baryon and Lepton quantum numbers
is not feasible. Yet for strong interactions, which respect strangeness and
isospin, the possible existence of charged meson balls of strangeness
and/or isospin as resonances in the low energy spectrum of QCD
has been considered \cite{DHS,A} as a possibility.

Recently it was pointed out 
\cite{DKSW}
that non-topological solitons generically appear in the
Minimal Supersymmetric Standard Model (MSSM).
More generally, supergravity (SUGRA) induced
logarithmic corrections in the scalar potential, as
well as non-renormalizable polynomial interactions that
appear naturally in the flat directions of the MSSM 
\cite{DRT}, 
give rise to baryon and lepton balls \cite{EM}.
They are composed of squarks
and sleptons and are very efficient ``repositories''
of baryonic and/or leptonic charge respectively. 
We will henceforth call them Q-superballs and denote them as Q-sballs.
They convert ordinary fermionic matter carrying a net
$B$ and/or $L$ charge into its bosonic counterpart.
In cosmology, large $B$ and $L$-balls can be generated from
decaying Affleck-Dine 
\cite{AD} 
condensates
that develop typically in the aftermath of an inflating
SUSY phase. In an expanding universe,
a coherently oscillating AD condensate  with a net
baryon charge is unstable to space dependent
perturbations decaying into
large baryon sballs
\cite{DKSW}.
$B$ and $L$-sballs, if unstable and rapidly decaying, could have contributed
to the net baryon number of the universe 
\cite{EM}.
If they are metastable but sufficiently long
lived till the present, they can be a
component of the much sought after cold dark
matter.
Non-abelian Q-sballs have been also discussed in
Wess-Zumino models\cite{AFK}. They minimally arise in renormalizable scalar potentials
with cubic interactions that respect supersymmetry and constitute domains that break it explicitly.

In the present paper we take a superstring inspired view on the
Q-sball bearing flat potentials in supersymmetric extensions of the standard model.
We ``embed'' the $U(1)$ ball bearing
MSSM flat directions in the ``effective''
low energy superstring picture.
We do it by considering low energy string  no scale effective
lagrangians such as the flipped $SU(5) \times U(1)$
\cite{ANT} 
model.      

We establish a precise mapping between the
low energy operators of different dimensionalities, such
as the $(Q\ell d^c)^2, u^cu^cd^c$ and $(u^cd^cd^c)^2$, and their
high energy operators they correspond to in these models.
Conceivably they are associated with the
small distance $Q$-sball bearing superstring induced flat directions.
We further address the question of proton stability in
conjunction with these operators and determine the conditions
to avoid fast proton decay. 

In the more general context of the effective lagrangians
we are considering, we generically  identify
an explicit $U(1)$ breaking term.
We consider its effect on the stability of
the $B$-sballs in the low energy regime ($E < M_s \equiv 1TeV$, 
the SUSY breaking scale). 

Finally we address the possibility that Q-sballs 
are present in the ``hidden''
sectors of supersymmetric theories which are a generic
 feature of supergravity
and more generally superstring theories. Shadow Matter has been a subject
of intense scrutiny recently with regard to its possible rich astrophysical and
cosmological implications\cite{KST}.

The paper is organized as follows:
In sections 2 we review $Q$-ball bearing flat directions in
the MSSM and consider the most general form of the
superstring inspired scalar potential with its one loop contribution
which is presented in section 3.    
In section 4 we identify the leading small distance
operators of $d=4$ and $d=6$ which correspond to the
flat directions  in the flipped $SU(5) \times U(1)$ model.
Large scalar vevs in those directions with a
nonzero baryon number generate the AD type of condensates
which can decay into $B$ and $L$-sballs.
We present precise results using renormalization
group results for the small-large distance evolution
of $U(1)$-sball configurations.  

\section{ Abelian Q-Sballs from Flat Directions}

In a scalar field theory with a global continuous symmetry, 
Q-balls
appear if the minimum of the quantity
$\left[2{\cal V}/\phi^2\right]$ occurs at some point
$\phi_0\neq 0$, where ${\cal V}$ is
the potential and $\phi$ is the scalar field.

In supersymmetric theories, $Q$-sballs are associated  with  the $F$-
and $D$-flat directions of the superpotential.  
In general, the flat directions (usually called the moduli space) are 
parametrized by expectation values of massless chiral fields (moduli), 
while along these directions the scalar potential vanishes.  In other 
words, supersymmetric theories have no classical potential along  
flat directions. 
In realistic supersymmetric theories, the role of
the fields acquiring vacuum expectation values (vevs) is played by 
particular combinations of the scalar quarks and leptons. 
The potential along these directions appears as a result of supersymmetry
breaking, radiative corrections  and non-renormalizable terms.
In MSSM, we are interested in forming operators invariant
under the gauge group (and therefore acceptable to appear
\footnote{ Additional discrete symmetries and string selection rules may 
prevent the appearance of othewise gauge invariant terms in the
superpotential.}
in the superpotential). Such operators can be formed by 
considering single field flat directions in a gauge invariant way.

As an example, we start with
a flat direction in the MSSM, by considering the operator
\beq
{\cal X}= Q_1\ell_1 d_2^c
\eeq
where indices denore generations.
The operator ${\cal X}$ has  $B-L=-1$. To see why it corresponds to a flat
direction of the superpotential, we consider all relevant Yukawa  terms
\bea
{\cal W} &=& \la_u^{11} Q_1 H_u u_1^c + \la_d^{11} Q_1 H_d d_1^c
         +  \la_d^{22} Q_2 H_d d_2^c +\la_e^{11}\ell_1H_ee_1^c+\cdots
         \label{sup1}
\eea
where dots stand for terms not involving the fields $Q_1,\ell_1,d_2^c$. 
The scalar component of
such operators could be parametrized by  a single scalar field $\phi$. 
{} For example, in the case of the
operator ${\cal X}$, by writing
\beq
Q=\phi\sin\xi,\quad
L=\phi\cos\xi\sin\theta, \quad
d^c=\phi\cos\xi\cos\theta
\eeq
the operator can be written
\beq
{\cal X}=(\sin\xi \cos^2\xi \sin\theta \cos\theta)\phi ^3\equiv c\phi^3
\eeq
It can be easily checked that ${\cal W}$ is $F$-flat with respect to the
derivatives of all the fields 
appearing in the superpotential.
This means that for
$\langle \phi \rangle \not= 0$, all derivatives  
${\partial {\cal W}}/{\partial f_a} = 0$, 
where $f_a$ stands for all fields.

The connection between Q-sballs and flat directions comes in when addition
of new terms alter these flatness and the potential takes the desired form
to accommodate Q-sballs. By giving non-zero
expectation values to these fields, the gauge group breaks down partially or
even totally (in the example of the operator ${\cal X}$, the standard model
group is broken down to $SU(2)_{colour}\times U(1)$). Also, since the
fields carry baryon $B$ and lepton $L$ numbers, the operator may also have 
a non-zero $B$, $L$ or $B-L$ quantum number. 

Let us now be more specific on the ways that a flat direction can be lifted.
This can be achieved 
by supersymmetry breaking effects, by non-renormalizable terms that could appear 
in the superpotential and finally by one loop corrections to the 
supersymmetry breaking induced soft mass terms.
The tree-level part of the scalar potential has the following types of terms: 
\bea
{\cal V}_{eff}&=&\mid\frac{\partial{\cal W}}{\partial f_a}\mid^2
           +A ({\cal W} + {\cal W^*}) +
            B (\frac{\partial{\cal W}}{\partial f_a}+c.c.)+
             \tilde{m}^2_{a}|f_a|^2
           \label{Veff}
\eea
The last three terms are  related to the supersymmetry breaking effects.
$A$ and $B$ are soft parameters of the order of the supersymmetry scale
while $\tilde{m}_a$ are the soft masses of the scalar components of the
superfields.

Let us see now how the non-renormalizable terms can lift the flatness of 
the potential (one loop effects will be discussed in the next section).
Suppose that a certain operator, describing a flat direction
and parametrized by the scalar $\phi$, takes
the form $\phi^m$; $m$ is a power which shows the number of fields 
entering the operator (in our previous example of the operator ${\cal X}$,
$m=3$). The operator cannot appear
by itself in ${\cal W}$, however, it can show up through NR-terms
together with other fields of the theory. Suppose, $f_x$ is such a field,
then a term of the type described above will have the form
\bea
{\cal W}&=&\la_{nr} f_x \frac{\phi^{m\cdot k}}{M^{m\cdot k-2}}\nonumber\\
         &=&\la_{nr} f_x \frac{\phi^{n-1}}{M^{ n-3}}
         \label{sup2}
\eea
where $k$ shows the power that the operator appears and $M$ is
some high scale. Obviously, the derivate with respect to the field $f_x$
leaves a non-zero term
\bea
\frac{\partial {\cal W}}{\partial f_x}\propto
    | \langle \phi \rangle |^{n-1}\not=0
\label{dsup}
\eea
and the flat direction is lifted. 
In addition, there are $A$-terms of the form $A \phi^n/M^{n-3}$. There are
two important points worth mentioning: 
First, 
since a  flat direction is parametrized by a single (scalar) field,
the soft supersymmetry breaking $A$-term in the potential
violates any possible $U(1)$ that the superpotential might respect.
The absence of a continuous $U(1)$ forbids  the
appearance of stable  $Q$-sball like solutions.
Thus, the question  arises whether there exist certain conditions 
such that  $Q$-sballs can be formed in the MSSM potential.
An apparent way out would be to require the initial condition for
the $A$-term to be zero, however 
renormalization group running effects will drive its value to
magnitudes comparable with other soft parameters. Nevertheless
the $A$-term could be kept relatively small  in the
interesting energy region, allowing the possibility for an 
unstable $Q$-sball  to develop.
Second,
in the case that a Q-sball can be formed, the sign of the $A$ parameter
plays evidently a crucial role in developing the required minimum of the
quantity ${\cal V}/|\phi|^2$.

Returning to the scalar potential, (\ref{Veff}) 
assumes the following general form
\bea
{\cal V}_{0}&=& {m}^2_S|\phi|^2+|\la_{nr}|^2
 \frac{|\phi |^{2(n-1)}}{M^{2(n-3)}}+ (A
 \frac{\phi^n}{M^{(n-3)}}+ c.c.)\nonumber
\eea
where $c.c.$ means complex conjugate and
 we assume for simplicity $A$ to be real
valued.  
We observe indeed that the continuous
$U(1)$ symmetry respected by the first two terms, is broken
by the $A$-term .  Clearly, the 
usual Q-sball solution $\phi = e^{\imath \omega t}\phi_0$ 
is no longer a solution of the equations of motion,
however, it can only be approximate one as long as the $A$-term is
relatively small.  

In our analysis, we
assume for simplicity only one scalar $\phi$. In the case of
multiple flat directions there will be more scalars, one for each 
such direction. The soft mass term ${m}_S$ is here related to the     
soft masses of the fields  making up the composite operator.  Its value is  
scale dependent $m(Q)$ and is calculated at any scale using the  
RGEs once the  initial value is known. At tree level, this mass is independent
of the scalar vev $\langle\phi\rangle$ and a minimum of the potential
at $\langle\phi\rangle=0$ is unavoidable.
 However, when one-loop corrections are
taken into account, there is a $\phi$-dependence of $m_S^2=m_S^2(\phi)$
which could possibly lead to a minimum away from zero. 

Thus, having defined a certain flat direction,
 the next task is the determination 
of the expectation value of the corresponding scalar parametrizing
this direction.  As stressed above, the tree-level classical potential
leaves the scalar vev undetermined.
 Radiative corrections to the classical potential
will determine this vev. Therefore, one has to add also at least one-loop
corrections to ${\cal V}_0$. 
The directions used to form condensates, should be chosen with great
care. The reason is that there are
$R$-parity breaking terms (like $d^c d^c u^c$)
which  create proton decay at low energies. We will work out cases
where fast proton decay is forbidden.

\section{ A Superstring Inspired Q-Sball Bearing Flat Potential }
A necessary presupposition to obtain a global minimum of 
${\cal V}_0/\phi_0^2$
away from $\langle\phi_0\rangle =0$, as can be seen from the form of
the effective scalar potential, is to have a $\phi$-dependent 
scalar mass parameter $m_S^2$. At the tree level, this is a sum of
soft mass parameters related to the fields forming up the condensate,
independent of the value of $\phi$. Thus, at tree level, 
${\cal V}_0/\phi_0^2$ has a minimum at
$\langle\phi_0\rangle = 0$. At the one loop
level the soft mass parameter $m_S^2$ is modified by one-loop corrections
proportional to the logarithm $\log{\phi_0^2}/Q^2$ where $Q$ is the
running scale. Thus, $m_S^2$ depends on $\phi$ and a minimum away from the
origin is possible. For a scalar field $\phi$, the
one loop corrected potential is 
\begin{equation}
\label{v1}
{\cal V}_1(\phi)={\cal V}_0(\phi) + \frac 1{64\pi ^2}
{m_S'}^2\left( \ln \frac{\phi ^2}{Q^2} -\frac32 \right)
\end{equation}
$V_0(Q)$ is the (R.G.E. improved) tree-level potential while the appearance
of the last term is due to the radiative corrections (at one-loop level).
Thus, in the case of a scalar field $\phi$, as that described above
the one loop correction results to a shift of the soft mass
parameter of the condansate. This makes the mass parameter
of the last term in (6) a function of $\phi$, which is essential
in the determination of the minimum of (1) at values $\langle
\phi\rangle \not= 0$ as required. The soft mass parameter will
be in general a linear combination of the scalar mass terms
forming the condensate, $m_S^2 =\sum_i \alpha_i m_i^2$. The general
form of the mass coefficient of the logarithmic term is then
\cite{KLNQ}
\bea
{m_S}^{'2}&=&\sum_i \alpha_i \frac{d\,\tilde{m}_i^2}{d t}\nonumber\\
     &=&\sum_i \alpha_i\left(-\sum c_A^ig_A^2m_A^2+c_t^i\lambda_t^2
         \sum_{n_3}\tilde{m}^2_{n_3}\right)
\eea
where $t=\log{Q}$ while $\alpha_i$ are coefficients.
{}Furthermore, $m_A$ are the gaugini masses and $g_A$ are the three
gauge couplings; further,
$$\sum m_{n_3}= \tilde{m}_{Q_3}^2+\tilde{m}_{u_3^c}^2
+ m_{\bar{h}}^2$$ 
is the sum of the scalar mass parameters of the  third
generation and the higgs while only 
$\lambda_t$ Yukawa contributions have been included.
Let's assume the particular case of $n=3$ which will be useful
in our subsequent analysis. In this case,  
the general form of the quantity ${\cal V}/\phi^2$ becomes
\bea
\frac{\cal V}{\phi^2} = \kappa +\nu \log\phi + \alpha \phi +
\lambda \phi^2
\eea
Since in our case we  take always $\lambda > 0$, 
it can be  checked that  the necessary 
minima with respect to $\phi$ exist for the cases 
$\nu <0,\; \alpha > 0$, and $\nu > 0,\; \alpha <0$.
Clearly, $\kappa, \nu, \alpha$ and $\lambda$ are scale dependent.
Their relation with the MSSM mass parameters $m_S^2$ etc are easily
found. 
To find the minima therefore, one has to examine the values
of the above quantities at any scale while varying the 
coefficients $\alpha_i$ in such a way so  as the 
above conditions are met.

In Fig.1 we plot the logarithm ${\cal V}_{eff}/\phi_0^2$
against the logarithm of $\phi_0$, at the energy scale
$Q=10^{13}GeV$. We have considered in the ${\cal V}_{eff}$ the zero
and one loop order potential (\ref{v1}) plus  NR- and
trilinear $A$-terms of the
form
\[  \lambda_{eff}\phi^6/M^2 + A_{eff}\phi^3
\]
We give to $\lambda_{eff}$ the value ${\cal O}(0.1)$
 while the three curves correspond to negative, zero and positive
 $A_{eff}$-values starting from the one that produces the deeper
minimum.  Yukawa effects have been included only due to
top quark.
(For presentation purposes, in the vertical axis
 an arbitrary positive constant has been added  to
${\cal V}_{eff}/\phi_0^2$.)
The minimum exists only when $A_{eff}$ obtains  negative values
($\sim -0.1 m_{3/2}$) in a narrow range, while it shifs to unacceptably high
values  as $A_{eff}$  changes due to renormalization group running.

\section{An $SU(5)$ superstring  model}
A natural ground for the above ideas is offered by models which
possess extra $U(1)$ symmetries. This is because such symmetries
(if properly chosen) may prevent disasterous  combinations of
$R$-parity breaking
terms which lead to fast proton decay. Models with these properties
are found in string constructions. As an example, we will work out
the relevant non-renormalizable operators which are obtained in the
case of the string derived flipped $SU(5)$ model. The details can
be found elsewere\cite{ANT,ELLN}.
 Here we recall only the necessary parts.
The generations and higgses are accommodated in
\bea
F=(10,-1/2) ,\; \bar{f} = (\bar 5,3/2),\;  \ell^c = (1,-5/2)\label{ff} \\
H=(10,-1/2),\; \bar{H}=(10,1/2);\;  h=(5,1),\; \bar{h}=(\bar{5},-1)
\label{hh}
\eea
The quark and lepton fields are found in the following representations
\bea
F = (Q, d^c, \nu^c),\; \bar{f}= (u^c,\ell),\; \ell^c = e^c
\eea
and the tree level superpotential relevant to the above
terms is
\bea
F F h + F\bar{f}\bar{h} + H H h + \bar H \bar H \bar h
\eea
Additional $U(1)$ symmetries can be chosen to distinguish 
between the various generations that appear at this level.
However, operators of the form described above, allowed by these
symmetries, can be generated in the non-renormalizable
 part of the superpotential. Then,
terms of the form (\ref{sup2}) appear in the effective potential of the MSSM
after the breaking of the GUT symmetry. We will concentrate in the case
of lowest dimension operators. Let us make the above by describing an
example. Suppose we are interested in the operator ${\cal X}=u^cd^cd^c$
describing a flat direction of the MSSM and we would like to check whether 
it could appear through the flipped $SU(5)$ non-renormalizable term
\beq
FFF{\bar f}\Phi
\eeq
 Here,  $\Phi$ is a possible singlet (or a power of singlet
fields) which may appear in such terms. The fields $F,\bar{f}$
are of the form (\ref{ff}) which may accommodate the ordinary
quarks and leptons, or other heavy fields of the same quantum
numbers if the model is non-minimal. The above NR-term gives the
following two low energy operators:
\beq
\begin{array}{lccccc}
&F  &  F  &  F  &  \bar{f}  &  \Phi\\
&\downarrow & \downarrow &\downarrow &\downarrow &\downarrow \\
&d^c(\bar 3,1)&d^c(\bar 3,1)&(1,1)&u^c(\bar 3,1)&(1,1)\\
{\rm and}& &&&&\\
&Q  &(1,1)& d^c &  \ell     &  (1,1)
\end{array}
\eeq
where the numbers in parentheses are with respect to the
$SU(3)\times SU(2)$.
When we take the derivative of this term with respect to 
the singlet $(1,1)$ belonging to $F$, a term of the form                       
\beq
\mid u^cd^cd^c\Phi\mid^2
\eeq
will appear in the effective potential. Likewise, a similar term
corresponding to the $QLd^c$ flat direction can appear through the same
fifth order non renormalizable term, again taking the derivative with
respect to the $(1,1)$ component of a $F$ field, the only one which can
be used to form a MSSM singlet term in the superpotential.
 
The $LLe^c$ MSSM flat direction can appear through the term
$F\bar f\bar f l^c \Phi$ where again the differentiation is taken with
respect to the  singlet of the $F$ field.
\bea
F\bar f\bar f l^c \Phi\ra LLe^c
\eea
These three directions,
namely $u^cd^cd^c$, $QLd^c$ and $LLe^c$, exhaust the 3-field composite
operators desribing flat directions in the MSSM.

Going now to the 4-field operators of the MSSM describing flat directions,
namely $QQu^cd^c$, $QQQL$, $QLu^ce^c$ and $u^cu^cd^ce^c$, its easily
checked that:
\begin{itemize}
\item
$QQu^cd^c$ and $QQQL$ can appear either from the $FFF\bar f\Phi$ operators
(differentiating with respect to $\Phi$) and from the
$FFF\bar f\phi_\pm\Phi$ ones (differentiating with respect to $\phi_\pm$).
\item
$QLu^ce^c$ and $u^cu^cd^ce^c$ can appear through the 
$F\bar f\bar f l^c\Phi$
operators (differentiating with respect to $\Phi$).
\end{itemize}
Higher order terms can lead to the same type of oparators with
some additional suppression factors that make such contributions
less important. It is therefore, adequate to  find
the minimum dimension NR-terms which contribute to a certain
type of operator.

A basic problem  encountered with this type of operators, however,
is the undesirable fast proton decay.
 In particular, the simultaneous existence of
terms as $u^cd^cd^c$ and $Q\ell d^c$ in the low energy effective
theory will induce a fast decay of the proton. 
             Thus, at first sight, it seems that terms forming
codensates for Q-balls should be baned due to their possible
catastrophic consequences and contradiction with the low energy
data. There are certain conditions, however, under which these terms 
can exist
without causing the aforementioned problems. In particular:
\begin{itemize}
\item
If the field $\Phi$ has a vanishing vev, $\langle\Phi\rangle=0$,
this operator cannot contribute to proton decay. However, the 
corresponding condensate survives in the scalar potential
when differentiating with respect to $\Phi$.
\item
If, as in the case of non-minimal models (which is often the
case in string constructions), one of the fields $F,\bar{f}$ entering
the operator is a heavy state, not related to the ordinary
quarks and leptons, proton decay is avoided. 
\end{itemize} 
Although the above requirements look rather unlikely to be fulfilled,
it is interesting that they do occur in certain string
models. In the following, we will investigate this possibility
in the case of the flipped $SU(5)$ string model. We will not
exhaust all possible cases here, but we will concentrate in
a particular operator.

To avoid confusion, we remark here that in
 the following, the indices in the representations $F_i,\bar{f}_i,\ell^c_i$ 
indicate the sector of the string basis they belong to rather than the generation.
In fact, the accommodation of the three generations takes place as follows:
\bea
\bar{f}_1 : u^c, \tau, \; \; \;
\bar{f}_2 : c^c, e/ \mu, \; \; \;
\bar{f}_5 : t^c \mu / e \nonumber \\
F_2 : Q_2, s^c, \; \; \;
F_3 : Q_1, d^c, \; \; \;
F_4 : Q_3, b^c \nonumber \\
\ell^c_1 : \tau^c, \; \; \;
\ell^c_2 : e^c, \; \; \;
\ell^c_5 : \mu^c
\label{assignments}
\eea                                                                           
where $F_i=(10,-1/2)$, $\bar{f}_i=(\bar 5,-1)$ and $\ell^c_i=(1,-5/2)$.
They also carry charges under the four surplus $U(1)$ factors which
will play a crusial role in determining the NR-terms.

Tree level couplings of the above model, do not lift flat
directions of the ones discussed above. There are fifth and sixth order
terms of this type which may lift the above flatness of quark
and lepton fields which might lead to fast proton decay. These are
\cite{ELLN}
\bea
F_4F_4F_3\bar{f}_3\Phi_{31},& F_2F_2F_3\bar{f}_3 \bar\Phi_{23},
 & F_1F_1F_3\bar{f}_3\Phi_{31}\\
F_3\bar{f}_3\bar{f}_1\ell^c_1\Phi_{31},&
 F_3\bar{f}_3\bar{f}_5\ell^c_5\bar{\Phi}_{23},&F_3\bar{f}_3\bar{f}_2\ell^c_2\bar
\Phi_{23}\\
F_3\bar{f}_2\bar{f}_2\ell^c_3\Phi_{31},&
 F_3\bar{f}_1\bar{f}_1 \ell^c_3\Phi_{31},& F_3f_5\bar{f}_5\ell^c_3\bar\Phi_{23}
\eea
However, with a proper choice of vacuum expectation values (\cite{ELLN})
which also  respects $F$ and $D$ flat directions, none of
these terms are dangerous since they do not involve particles
in the light Standard Model part of the spectrum.
At sixth order, the following potentially-dangerous operators appear:
\bea
F_4F_3F_3 \phi_+\bar f_5\bar\Phi_{23},&F_4F_3F_3\bar f_5 \bar\phi_-\Phi_{31}
\eea
The singlet fields $\phi_+$ and $\bar\phi_-$ do not acquire    
vevs and proton decay is avoided.
Being safe from proton decay problems, we turn now to the possible
role of these terms on the $Q$-ball formation. This will be manifest
in the way described for the  term of the form (\ref{sup2}). Thus 
in a number of cases, for example, the role of
the field $f_x$ is played here by the field $\bar{f}_3$ which appears
in a number of terms of fifth order. Differentiating with respect
to this field, we may create the quantity analogous to that in (\ref{dsup}).
(In fact, here, the operator is multiplied by one additional
singlet vev, namely $\langle\Phi_3\rangle$, which will offer an
additional supression factor to the NR-coupling:
$\lambda_{nr}\sim g_U \frac{\langle\Phi_3\rangle}{M_U}$.)
The preceding discussion does not intend
 to systematically exhaust
all possible sources accosiated with $Q$-ball formation. Rather,
it is indicative in the way these finite energy configurations may 
occur in realistic theories.
 The question on the stability of these objects and
a  thorough study  of the related equations  should be a first priority
before a complete analysis in the context of
such theories is done.
 
\section{Q-Sballs in the Shadow World - Concluding Remarks}

A Hidden Sector appears to be a generic element in supergravity theories and 
the low energy limit of any superstring theory.   
In addition to providing a gravity mediated mechanism to break supersymmetry
its possible role as being the origin of a purely gravitating matter component in our universe
such
as dark matter has been previously discussed\cite{KST}. 
As so far our discussion of Q-sballs appearing
 in the spectrum of  susy gauge theories
with unbroken global symmetries concerned 
the observable sector from the point of view
of the superstring we would now like to deal
 with Q-sballs appearing in the hidden
sector of any such theory. 
In the context of superstring inspired models
 a hidden sector typically contains scalar
particles with only gravitational
 interactions which are described to a very good approximation
by sigma models\cite{WB}.
 These are parametrized in general by a coset  $ G/H$ space. The group
$G$ acts nonlinearly whereas $H$ acts linearly
 and can be viewed as a global symmetry of the $\sigma$-model.
Hence we would expect the presence of
 abelian or nonabelian global symmetries and hence of
Q-sballs in the hidden sectors of such theories\cite{K}.   
In our present model such a sector has an $SU(4) \times SO(10)$ gauge symmetry. As such 
the $SU(4)$ coupling may become strong 
at around $\sim 10^{10-12} Gev$ mimicking pretty much QCD.
The confinement of the nonabelian gauge 
charge greatly restricts the meaninglfulness of extended
Q-sball configurations with a net 
nonabelian charge. In a more general setting, however, in the
presence of unbroken global symmetries 
the hidden sector should be expected to possess 
nontopological solitons which are solutions 
to the field equations of motion including gravity.

An interesting realization of this possibility was recently put forward
for the minimal non-susy
 electroweak theory\cite{new}. Q-Balls are shown 
to be present and induced by the
coupling
 of the unobservable so far Higgs to a gauge 
singlet complex scalar field in
 a theory with an additional unbroken global abelian symmetry.
 As these stable solitons presumably interact 
gravitationally and only through a
 Higgs exchange to the observable sector they can be a dark matter
 component.   

 In summary  we investigated the possible existence of Abelian 
non-topological solitons  associated with global $B$, $L$, or $B-L$ 
quantum numbers in low energy effective supergravity
models arising from superstring theories.   We described the conditions 
in the effective potential for a $B$-sball to appear and discussed the role
of radiative corrections, $A$-terms and non-renormalizable contributions.
In particular, we found that $Q$-sballs are likely to appear at high scales,
however, we showed that $A$-terms lead to a potential instability
of the associated $Q$-sball.  We further discussed the ways to ensure
proton stability triggered by the above non-renormalizable operators and  
presented a string example where all baryon and lepton violating terms
associated with these finite energy configurations are suppressed. 

\section{Acknowledgements}
We thank Alexandros Kehagias and Costas Kounnas 
for sharing their thoughts with us.

\newpage
\begin{figure}
\begin{center}
\hspace{-1cm}
\epsfig{figure=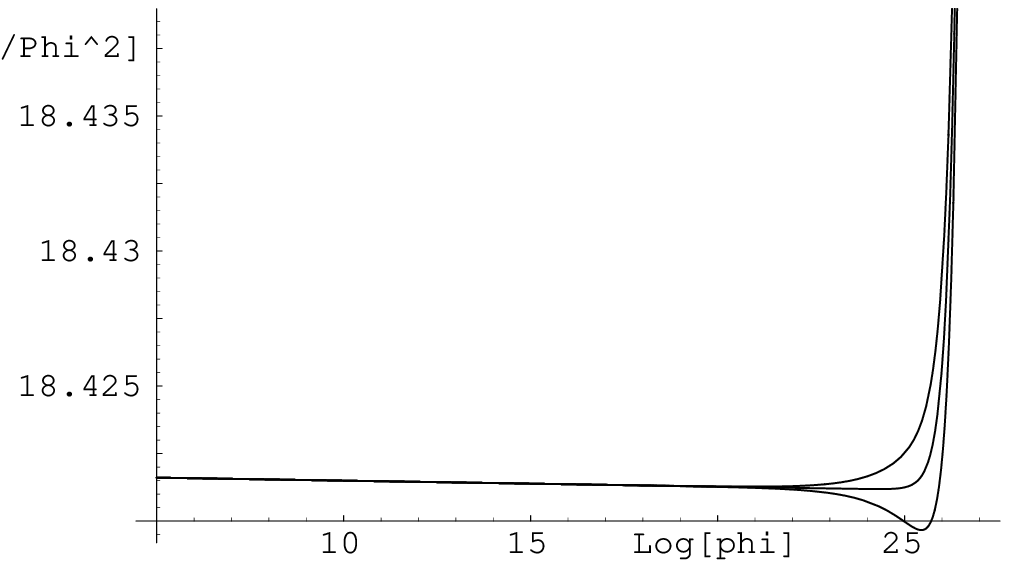,height=7truecm,width=10truecm}
\vspace{.5cm}
\caption{Plot of the quantity $Log[{\cal V}_{eff}/\phi^2]$
vs $Log[\phi_0]$ in the effective supergravity
model described in the text and for the operator $u^cd^cd^c$,
for three values of the parameter $A_{eff}$. The minimum is formed in a
very high scale, when the conditions  discussed in the text are met.
The corresponding superheavy $B$-ball is unstable.}
\end{center}
\label{fig:ampoff1}
\end{figure}       
\end{document}